\documentclass[twocolumn]{aa}
\usepackage{graphicx}
\begin{document}  
   \title{The optical and X-ray flickering of XTE~J1118+480}

   \author{J. Malzac\inst{1,2} \and T. Belloni\inst{2} \and H.C. Spruit\inst{3} \and G. Kanbach\inst{4}}

   \offprints{J. Malzac, malzac@ast.cam.ac.uk}

   \institute{Institute of Astronomy, Madingley Road, 
         Cambridge, CB3 0HA, United Kingdom
     	\and
         Osservatorio Astronomico di Brera,
              Via Brera 28, I-20121 Milano, Italy
         \and
             Max-Planck-Institut f\"ur Astrophysik, Postfach 1317, 85741 Garching, Germany
        \and
             Max-Planck-Institut f\"ur Extraterrestrische Physik, Postfach 1317, 85741 Garching, Germany}

   \date{Received ??; accepted ??}

   \abstract{
   
    We use both time-domain and Fourier techniques to study
   the correlated optical/X-rays variability in the black hole 
   X-ray nova XTE J1118+480. Its X-ray timing properties such as the
   shape of the X-ray power spectrum and 
   cross-correlation functions (CCFs) between X-ray energy bands, the
   slight decrease of RMS variability 
 from  30 \% in the 2-5.9 keV to 19 \% in the 15.5--44.4 keV band, 
    as well as the X-ray hardness/flux anti-correlation are very similar
    to what is found in other black hole binaries in the hard state.
 The optical/X-ray CCF 
  is virtually independent of the X-ray energies. The optical flux appears 
  to be correlated not only with the X-ray flux but also with the X-ray 
  spectral variability. Both the coherence function and the lags between
 optical and the X-rays are Fourier frequency dependent. 
   The optical/X-ray coherence function reaches its maximum
($\sim$0.3) in the 0.1-1 Hz range and the time-lags decrease 
 with frequency
 approximatively like $f^{-0.8}$. 
 The correlation between X-ray and optical light curves 
appears to have time-scale-invariant properties. 
The X-ray/optical CCF maintains a similar but rescaled shape on
time-scales ranging at least from 0.1 s to few 10 s.
Using the event  superposition method we show
that the correlation is not triggered by
 a single type of event (dip or flare) in the light curves.
 Instead, optical and X-ray  fluctuations of very different shapes, amplitudes
 and time-scales are correlated in a similar mode where  
  the optical light curve is seemingly
 related to the time derivative of the X-rays.

   }

   \maketitle

\section{Introduction}

%
%________________________________________________________________
   \begin{figure}
   \centering
   \scalebox{0.4}{\includegraphics{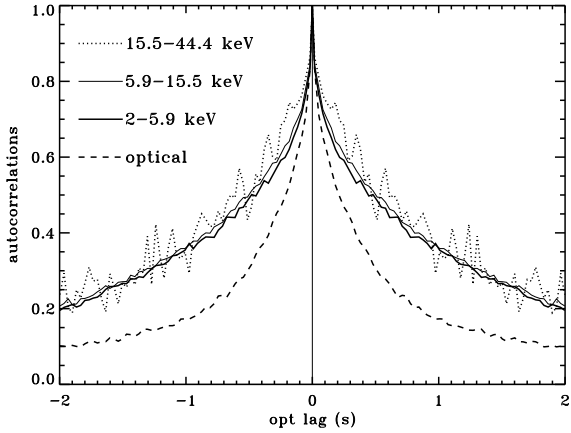}}

   \caption{Optical and X-ray ACFs of the photon flux in
the X-ray bands b1, b2, b3 and optical. In this figure as well as in
Figs.~\ref{fig:xxccfs} and \ref{fig:ccfox}, 
the low frequency noise on time-scales longer
than 100 s was 
removed by dividing the light curves by the piecewise linear trend.
This piecewise linear trend is defined as the linear interpolation
between the 100 s average count rate computed at time bins
separated  by 50 s.  
The presence of noise due to the
optical telescope oscillations (see SK02 and
Sect.~\ref{sec:coherlags}) significantly affects the shape of the ACF.
Therefore, only the 2 segments in which
 these oscillations are minimal were used to estimate the optical ACF.
For clarity, the ACFs were rebinned with 30 ms resolution.}
              \label{fig:optxccfs}
    \end{figure}

%______________________________________________ Gamma_1 (lg rho, lg e)
   \begin{figure}
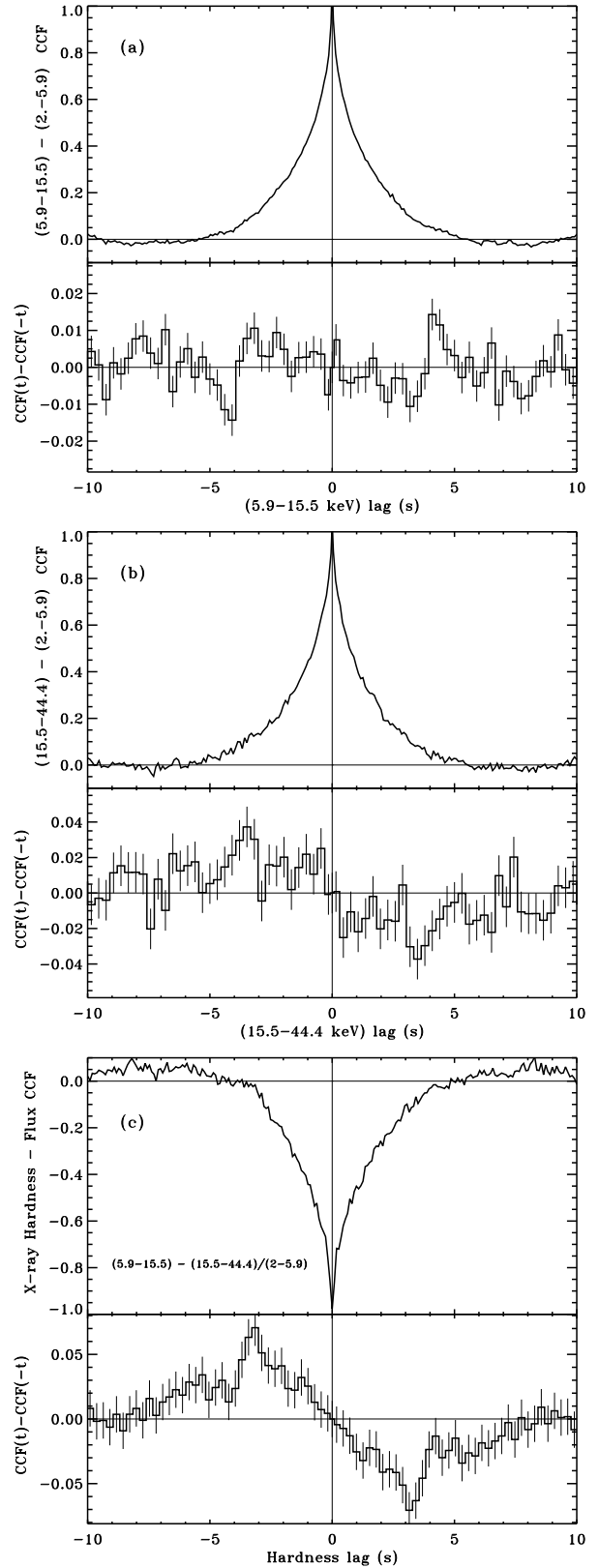

   \centering
   \scalebox{0.4}{\includegraphics{3498.f2a}}
   \scalebox{0.4}{\includegraphics{3498.f2b}}
   \scalebox{0.4}{\includegraphics{3498.f2c}}
\caption{X-ray CCFs and their anti-symmetric parts $CCF(t)$$-$$CCF(-t)$.  
{\bf a)} flux in band b2 vs flux in band b1. 
{\bf b)} flux in band b3 vs flux in band b1. 
{\bf c)} hardness ratio b3/b1 vs flux in band b2.
 The one sigma uncertainties on the anti-symmetrized CCFs were 
 estimated from the photon noise in the original data using the 
 standard statistical error propagation method.
The CCFs and anti-symmetrized CCFs were rebinned with 90 ms and 
300 ms resolution respectively}
\label{fig:xxccfs}
    \end{figure}

\begin{figure}
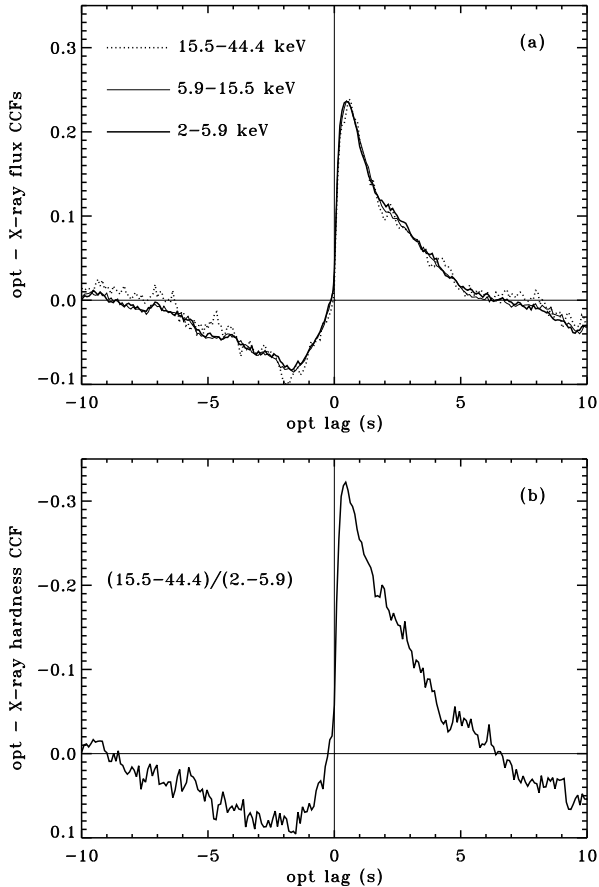

\centering 
\scalebox{0.4}{\includegraphics{3498.f3a}}
\scalebox{0.4}{\includegraphics{3498.f3b}}
\caption{Correlations between X-ray and the optical. 
 and X-ray flux. {\bf a)} X-ray flux vs optical flux CCF for the 3 X-ray bands. {\bf b)} X-ray hardness vs optical flux CCF.
For clarity the CCFs were rebinned with a 90 ms
resolution}
\label{fig:ccfox}
\end{figure}

\begin{figure}
\centering
   \includegraphics[width=\columnwidth]{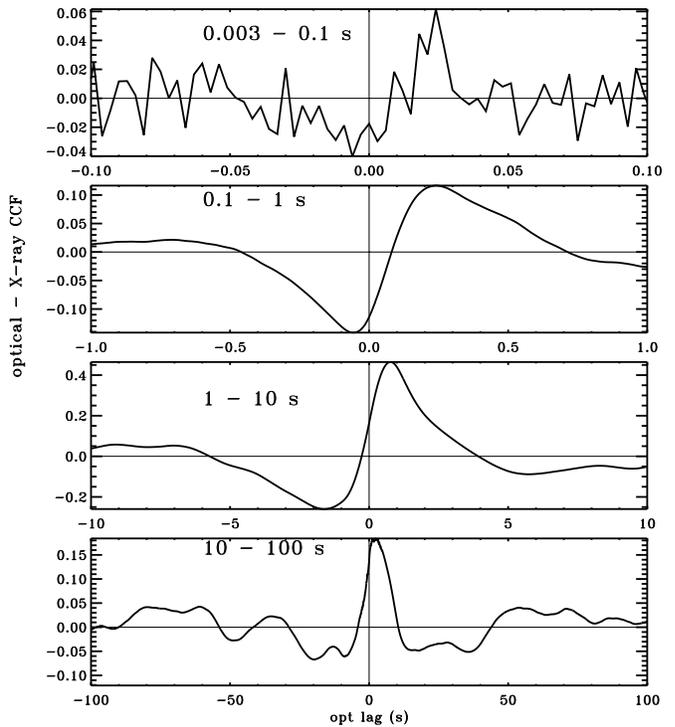}
\caption{Optical vs X-ray CCFs for different time-scales of the
fluctuations. The CCFs were computed after both light curve have been
filtered keeping only the time-scales in the range indicated. High
frequency noise was removed applying a box car filter, the low
frequency noise was removed by dividing the light curve by the
piecewise linear trend.}
         \label{fig:ccfscadep}
   \end{figure}

%
%________________________________________________________________
   \begin{figure}
   \centering
  \includegraphics[width=\columnwidth]{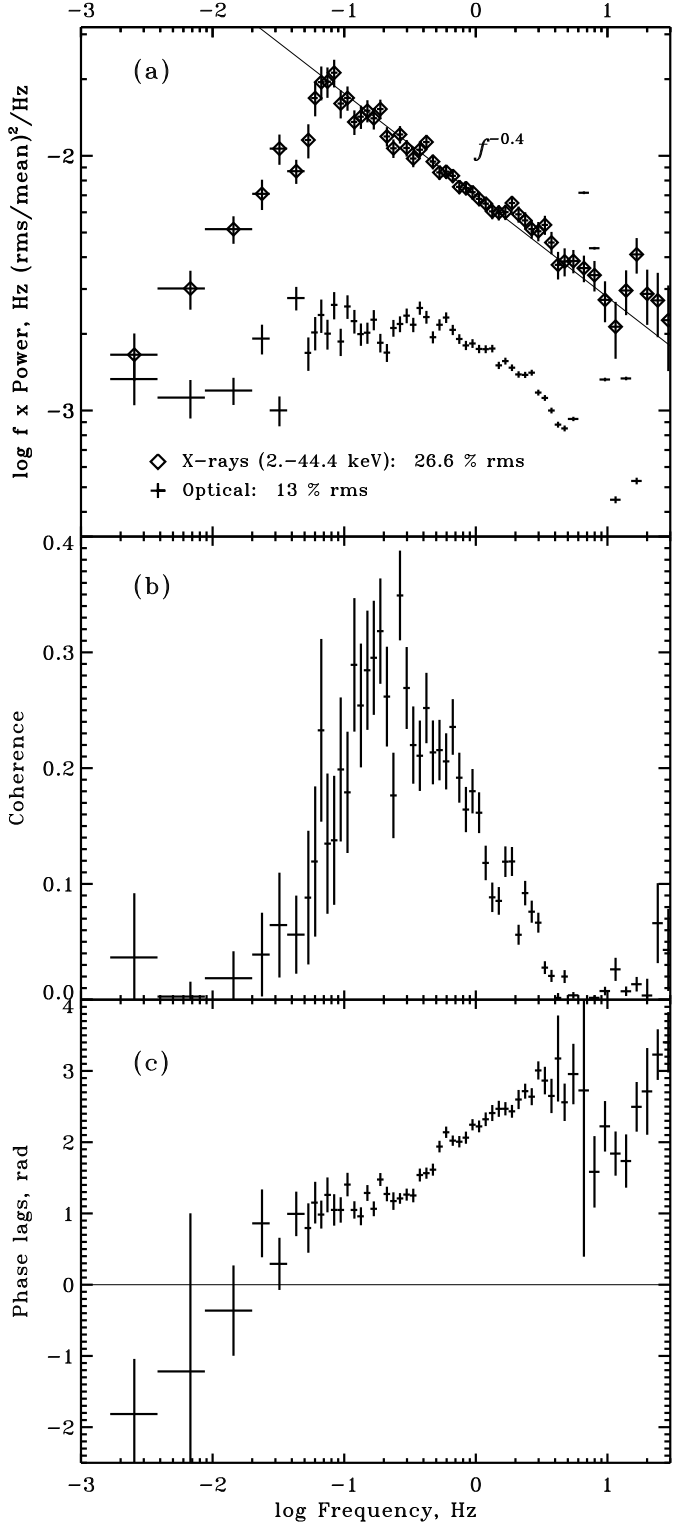}
   \caption{The optical/X-ray correlation in the Fourier domain.
 {\bf a)} X-ray and optical power spectra. The counting noise was
subtracted (see Sec.~\ref{sec:psd}).
 {\bf b)} X-ray/optical
coherence. {\bf c)} phase-lags as function of
Fourier frequency. A positive lag implies 
that the optical is delayed with respect to the X-rays. Uncertainties
where computed according to Eq.~8 of Vaughan \&
Nowak (\cite{vaughan1997}) for the
coherence, and Eq.~16 of Nowak et al. (\cite{nowak1999}) for the phase-lags. All
error bars are at the 1 sigma level.}

              \label{fig:corxofour}
    \end{figure}

%                                     Two column figure (place early!)
%______________________________________________ Gamma_1 (lg rho, lg e)
   \begin{figure}
   \centering
  \includegraphics[width=\columnwidth]{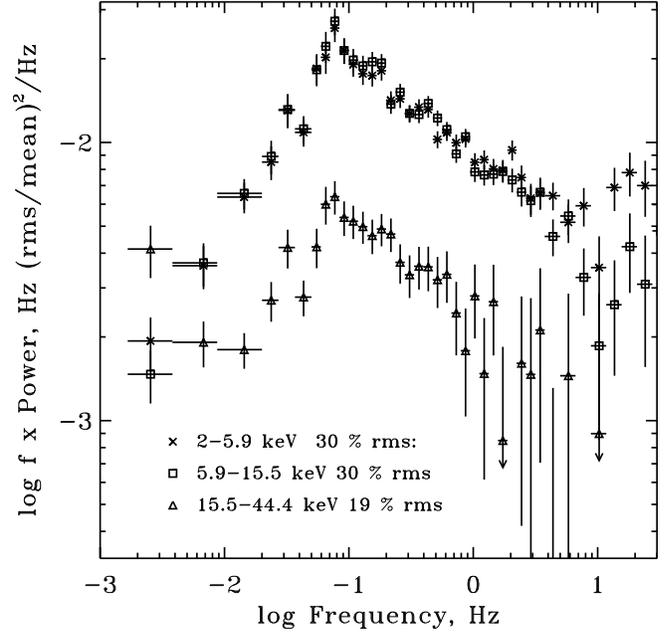}
   \caption{X-ray power spectra in bands b1, b2 and b3. 
 The counting noise was subtracted (see Sec.~\ref{sec:psd}).}
              \label{fig:pds}
    \end{figure}

The  X-ray nova XTE J1118+480 (=V$^{*}$ KV
 UMa), was discovered by the Rossi X-Ray Timing Explorer
({\it RXTE}) All-Sky Monitor ({\it ASM}) on 2000 March 29 
(Remillard et al. \cite{remillard}) as a slowly rising source.
 The outburst lasted a few months with a plateau-like
 X-ray light-curve that rapidly fell down  around mid-July. 
A post-analysis revealed
an an earlier outburst in January 2000. 
 The optical spectrophotometry proved a low mass X-ray binary system
 containing a black hole of at least 6  $M_{\sun}$
(Mc Clintock al. \cite{mcclintock2001a}, Wagner et al. \cite{wagner2001}).
 The interstellar extinction towards
 the source is exceptionally low (Garcia et al. \cite{garcia2000}).
 This fact allowed an unprecedented wavelength coverage
 (Mauche et al. \cite{mauche2000}; Hynes et al. \cite{hynes2000}; 
McClintock et al. \cite{mcclintock2001b};
 Hynes et al. \cite{hynes2003}; Chaty et al. \cite{chaty2003} and reference
 therein). In  the radio to optical bands,  
a strong non-thermal component was associated with synchrotron
emission from a powerful jet or outflow (Fender et
al. \cite{fender2001}). In the
optical to EUV bands the spectral energy distribution 
 is dominated by a thermal component from the accretion
disc. The X-ray emission consists in a typical powerlaw spectrum with photon
index $\Gamma \sim 1.8$. Such a  
spectrum is generally associated with Comptonisation in the hot inner  part
of the disc or corona. In the case of XTE~J1118+480,
other interpretations in terms of jet emission
 through inverse Compton (Georganopoulos, Aharonian \& Kirk,
\cite{georganopoulos2002})  or synchrotron (Markoff, Falcke \& Fender
\cite{markoff2001}) were also put forward.

The X-ray variability is characterized by a flaring activity on time-scales 
of a few seconds typical of accreting black hole sources
 (Revnivtsev et al. \cite{revnivtsev2000}). A $\sim0.1$ Hz
quasi-periodic oscillation (QPO) was reported
 by the same authors
 and subsequently confirmed by the {\it ASCA} data (Yamaoka et al.
 \cite{yamaoka2000}) 
as well as other {\it RXTE} observations where it appeared to be variable
 (Wood et al. \cite{wood2000}).
The simultaneous {\it RXTE/HST} observations showed that a similar QPO is 
also present in the {\it HST} data and that the X-rays and optical/UV bands are
 correlated on time-scales of seconds 
(Haswell et al. \cite{haswell2000}, Hynes et al. in preparation).

Kanbach et al. (\cite{kanbach2001}, hereafter K01) performed 4  
observations 
on the nights of 4, 5, 6,  and 7 July 2001, shortly before the end 
of the outburst.
They combined simultaneous observations with {\it RXTE} and the rapid optical 
photo-meter {\it OPTIMA} (Straubmeier et al. \cite{straubmeier2001}) attached to the 1.3 m telescope on Mt Skinakas, Crete. 
They studied the X-ray and optical auto-correlation functions
 (ACFs) and X-ray/optical cross-correlation (CCFs). 
They found the optical ACF to be significantly narrower than the
X-ray ACF. The full width at mid height (FWMH) of the two ACFs
 differs by factor $>2$. They found the correlation between the
  optical and X-ray light curves to be surprisingly complex.
The CCF rises very quickly at positive optical lags, peaks around $0.5$~s 
and then declines slowly at larger lags.
Therefore, the optical light curve tends to be delayed 
 relatively to the X-rays by a relatively long time (of order of 1 s). 
Strikingly, the two bands appear to be anti-correlated at negative optical
 lags indicating a systematic
 optical dip 1-2 s before the X-rays reach their maximum.
As this feature suggests that the optical band 'knows' about what will
happen in the X-rays about 1 second later, it was named
`precognition dip' by K01.
K01 further showed that the detailed shape of the CCF
 is variable on time-scales as short as 30 s.
Spruit \& Kanbach (2002, hereafter SK02) analyzed this variability of
the CCF in terms of a principal component analysis. They found 
that there are two systematic components which vary statistically 
independently. Both have the shape of a broad dip followed by a
sharper peak, but their duration differs by a factor of about 3.   

K01 argued that those characteristics rule out reprocessing as the 
origin of the optical variability and favors synchrotron emission.
 Merloni, Di Matteo and Fabian (\cite{merloni2000}) had previously 
interpreted the correlated X-ray/optical  variability in terms of 
thermal Comptonisation of synchrotron radiation in an accretion disc corona.
 It is not clear however how such a model could reproduce the complicated 
shape of the cross-correlation function.
K01 proposed a somewhat different
 scenario where  
the X-rays are produced within a few hundred Schwarzschild radii from the black hole and the optical
 is  produced by synchrotron effect
at much larger distance, in a magnetic outflow. In this framework  the optical lags
 would be due to the travel-time of a perturbation from the disc
 to the optical photo-sphere of the outflow. This scenario however (as well 
as reprocessing models)  does not explain the  `precognition dip'.

 The only other source for which 
simultaneous X-optical observations at millisecond time resolution
have been reported is GX 339-4, where an anti-correlation between the
X-rays and optical, similar to
the 'precognition' dip, was observed (Motch et al. \cite{motch1982}).
Fabian et al. (\cite{fabian1982}) interpreted
the optical emission as cyclo-synchrotron radiation produced 
in the inner parts of the accretion flow.

In this paper we use the data of K01 to investigate further the optical X-ray 
correlation in XTE J1118+480.
 The optical observations and data processing are described in K01
 and SK02. The PCA X-ray light curves used in these previous analyses were
 decomposed into 3 energy bands: 2--5.9 keV (referred hereafter as b1),
 5.9--15.5 keV (b2) and 15.5--44.4 keV (b3). 
In order to study a range of time-scales, 
we used only the data for which we had a sufficiently long continuous 
exposure. The analysis was performed over 
11 light curves segments of 590 s duration and 3 ms resolution
 taken
during the four observation nights. 
Most of the results shown in this paper 
are averaged over all these data segments.
The paper is divided into three sections presenting successively
the results from the time-domain, Fourier domain and superposition
analysis. These results are then summarized and briefly discussed in 
Sect.~\ref{sec:discussion}.

\section{Time domain analysis}\label{sec:tdanal}

\subsection{X-ray properties}\label{sec:tdxr}
%                                     Two column figureplace early!)
%______________________________________________ Gamma_1 (lg rho, lg e)

We first examine the properties of the X-ray light curves in the time
domain. 
The X-ray ACFs in the three energy bands are, within 
the uncertainties, indistinguishable from the total X-ray ACF 
and broader than the optical ACF (as shown in  
Fig.~\ref{fig:optxccfs}). The FWMH of the X-ray
CCFs is $\sim 1 s$ while that of the optical is only of $0.4$ s. 
The X-ray CCFs  of the different energy
 bands, shown in the panels (a) and (b) of Fig.~\ref{fig:xxccfs}, are all
strongly peaked 
at lags $<30$ ms with a shape very similar to that of the ACFs.
The different energy bands are thus strongly correlated at (nearly) 0 lag.
 These results are very similar to that of
Maccarone, Coppi \& Poutanen (\cite{maccarone2000}) for Cygnus X-1. 

 We then searched for correlations between variations of the X-ray hardness 
and the X-ray flux.
We computed the ratio  
of the two light curves in two energy bands b3 and b1. The resulting
time series represents the time dependent hardness b3/b1.
Then we computed its CCF with the third energy band b2. 
In practice, the 
hardness ratio b3/b1 cannot be computed for the b1 time bins containing 0
counts. We thus rebinned 
the light curves with a 90 ms resolution before estimating b3/b1,
 so that zero count bins are reduced to a negligible fraction 
of the total number of bins. For the few remaining bins with zero count,
b3/b1 was arbitrarily set to the time averaged hardness ratio.

Panel  of (c) of Fig.~\ref{fig:xxccfs}
 displays the resulting hardness vs flux CCF.
Clearly the count rate 
in the b2 band is strongly anti-correlated with hardness at (almost) zero lags.
 The source is thus softer when brighter for fluctuations of time-scale of
 seconds.
We note also that this CCF shape is similar to the shape of the X-ray ACFs,
 suggesting that spectral variations map very closely the flux.
We also computed the hardness-flux correlation for the different
 combinations between our three energy bands and found very similar results.
The b2/b1-b3 CCF appeared noisier indicating that the amplitude
 of the fluctuations of the b2/b1 hardness is lower than that of 
b3/b1 or b3/b2.
Such a hardness/flux anti-correlation looks similar to what found in Cygnus 
X-1 in the hard state (see e.g. Li et al \cite{li1999}; Feng et al.
 \cite{feng1999})

The antisymmetrized CCFs (i.e. CCF(t)-CCF(-t)), also
shown in Fig.~\ref{fig:xxccfs}, indicate that the X-ray CCFs as well as 
the hardness flux CCF are asymmetric for lags in the approximate range
1--5 s
 (outside this range the poor statistics in the XTE J1118+480 data did
 not enable us to firmly establish any asymmetry).
The shape of the asymmetries suggests that the spectrum is
relatively harder during the rising phase of a flare than its decay.
In term of lags between X-ray bands this implies that 
the soft band is delayed with respect to the hard one.
This kind of asymmetry is thus opposite to that found by
 Maccaronne, Coppi \& Poutanen (\cite{maccarone2000}) 
in the X-ray CCFs of Cygnus X-1, and more generally to  
the hard lags often reported in  hard-state sources.

\subsection{Optical/X-ray correlations}\label{sec:oxcor}

   \begin{figure*}
   \centering
   \scalebox{0.4}{\includegraphics{3498.f7a}}\hspace*{0.5cm}\scalebox{0.4}{\includegraphics{3498.f7b}}

   \caption{{\bf a)} X-ray coherence and {\bf b)} phase lags. 
Squares stand for b2 versus b1, triangles for b3 vs b1. Due to the
high noise level it has not been possible to estimate conveniently the b3 vs b1 
coherence in the higher frequency bin.}
              \label{fig:cophx}
    \end{figure*}  

%                                     Two column figure (place early!)
%______________________________________________ Gamma_1 (lg rho, lg e)
  \begin{figure}
   \centering
  \includegraphics[width=\columnwidth]{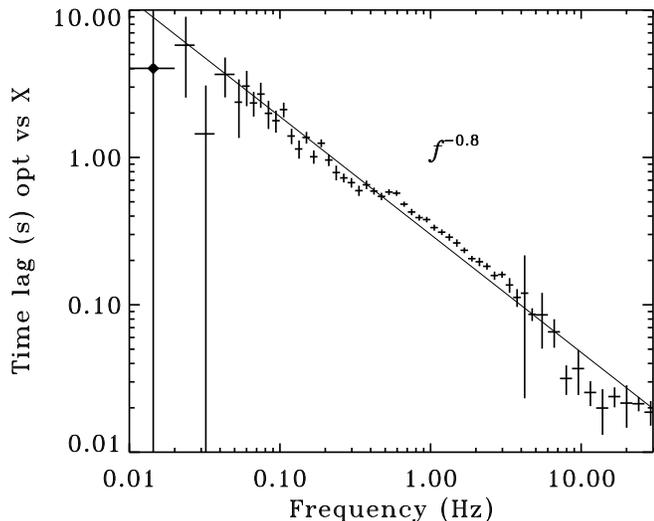}
   \caption{Frequency dependent optical time-lags. Filled diamonds stand for negative lags.}
              \label{fig:lago}
    \end{figure}

We then investigated the dependence of the optical/X-ray correlation
 on the X-ray energies. We found that the optical/X-ray CCF does not depend
 significantly on the X-ray band considered.
 Actually, our 3 optical/X-ray CCF are identical within the
uncertainties, as shown in Fig.~\ref{fig:ccfox}a.
 
The optical/X-ray \emph{hardness}   CCF (shown in Fig.~\ref{fig:ccfox}b) is
qualitatively similar to the optical/X-ray \emph{flux} CCFs
 modulo a mirror symmetry around the X-ray axis.
 The optical flux and X-ray hardness are indeed anti-correlated
at positive optical lags and positively correlated at negative lags.
This feature derives naturally from the X-ray flux hardness
 anti-correlation demonstrated in the previous section. 
It follows that the optical flux is correlated not only with
 the X-rays count rate but also with the X-ray spectral variability.
Similar results were found for the hardness ratios computed with
 the different combinations of the 3 bands. 

The time-domain analysis presented above
  provides only information (such as time lags) that is 
 averaged over all times-scales present in the signal.
However, in general, the properties of time-series  or the correlation between
 two time-series does depend on the time-scale considered.
In order to study the dependence of the optical/ X-ray CCF on
the time-scale of the fluctuations we thus filtered  
the light curves before computing the CCF  keeping only a small
range of time-scales in both signals.

The results are displayed in Fig.~\ref{fig:ccfscadep}. 
For all of the different time-scales considered,
 both the anti-correlation at negative lags and
positive correlation at positive lags is present.
The characteristics lags appears to depend on the time 
scale of the fluctuations.
The lag of the optical peak increases from $\sim$ 30 ms for time-scales
shorter than 0.1 s to $\sim$ 3 s for the 10--100 s fluctuations.
Similarly the optical dip ranges from $\sim$ -6 ms to $\sim$ -30 s, 
so that, in the rescaled plot, we see both the optical peak and dip shifting 
toward negative lags as the time-scale of the fluctuations increases. 
Still in the rescaled plot, 
the optical flare appears narrower at longer fluctuations.

Despite these significant differences, 
the overall shape of the CCF appears similar but rescaled over a
range covering more than three decades of time-scales.

\section{Fourier analysis}
 
In the following, we will use a Fourier analysis to 
explore more quantitatively the dependence
of the optical/X-ray correlation in XTE~J1118+480 
on the time-scale of the fluctuations.

\subsection{Power density spectra } \label{sec:psd}

Fig.~\ref{fig:corxofour}a  shows the X-ray and
optical power density spectra (PDS). The PDS were computed, averaged, rebinned
and normalized to the total rms variability 
as described in Nowak et al. \cite{nowak1999}. The counting noise was
approximated by a white noise component that was then
 subtracted from the total PDS.

The fractional rms variability amplitude integrated over the frequency 
range 10$^{-3}$--$10^{2}$ Hz is about 26 \% in the X-rays and 13 \% in the optical.
The optical and total X-ray power spectra were already shown 
 and discussed in SK02.
 Here we will just note the similarities of the X-ray PDS
 with that of other black hole candidates in the hard state such
 as the plateau like shape up to $\sim$ 0.1 Hz and then a power law component. 
We note also that in our July observation the X-ray PDS 
is slightly different from
 that found in the earlier observation of Revnivtsev et al. (\cite{revnivtsev2000}) with a 
break at higher frequencies ($0.1$  Hz instead of $0.03$ Hz) a flatter
 power-law component with slope $\sim 1.4$ as compared to the $1.6$ slope
 reported by Revnivtsev et al. \cite{revnivtsev2000}.
This evolution of the power spectrum was accompanied 
by a significant reduction of the total rms amplitude that
 was ~40 \% during the observation of Revnivtsev et al. 
 More strikingly,  the power spectra of Fig.~\ref{fig:corxofour}a
 show little to no evidence for the 
QPO reported in previous observations both in the X-ray and optical
(see e.g. for comparison Fig.~1 of Revnivtsev et al. \cite{revnivtsev2000}).
This is however consistent with the previously reported trend
 in the evolution of the timing properties of the source during 
the outburst. Indeed the {\it RXTE} monitoring campaign showed
 that all the characteristics 
frequencies increased  with time  and the QPO became much weaker
 (Wood et al.~\cite{wood2000}; Hynes et al. in prep).

As shown in Fig.~\ref{fig:pds}, the X-ray PDS depends
only slightly on energy. We find that the PDS in the two 
lower energy bands (b1 and b2) are extremely similar. On the other hand, 
at higher energy (b3) the shape of the PDS differs significantly at
low frequency and the integrated rms amplitude
 is lower (19~\% instead of $\sim$~30~\% in b1 and
b2). 
 A similar energy dependence of the rms variability amplitude
is reported in Cygnus X-1 in the hard state (e.g. Nowak et al. \cite{nowak1999}).

\subsection{Coherence and lags}\label{sec:coherlags}

The coherence function represents the degree of linear correlation between
 two time series as a function of Fourier frequency (Vaughan and Nowak
 \cite{vaughan1997}).
The coherence function between different X-ray bands appears to be unity
 over a wide range of frequencies in black hole candidates (Nowak et al.
 \cite{nowak1999}).
We found a similar behaviour in XTE~J1118+480 as shown in 
Fig.~\ref{fig:cophx}a.

The X-ray phase lags between energy bands are shown on Fig.~\ref{fig:cophx}b.
Due to the poor statistics of the data, the phase
lags are very difficult to measure.
In order to improve the statistic we averaged the
frequency bins over a broad range of frequencies (nearly  a decade).
Even so, we get essentially upper limits that are above the typical
amplitude of the hard lags reported in Cyg X-1 or GX 339-4 in the
hard state.
There is however an indication that the lags are increasing with
frequency,
and surprisingly, the lags tend to be negative at low frequency.
In particular in the frequency range 2 10$^{-3}$--2 10$^{-2}$ Hz,
 the phase lag between b3 and b1,
 is detected at the 3 sigma level and is negative.
This implies that the soft band leads the hard band by 2 to 20 s in this 
frequency range. This is qualitatively consistent
 with the asymmetry found in the X-ray CCFs (see Sect.~\ref{sec:tdxr}). 
This result should however be regarded with caution since it involves
frequencies that correspond to time scales approaching the
duration of the light-curve segments.
 In an attempt to improve the statistics, we used 
 public \emph{RXTE} archive data to compute the phase-lag spectra with 
a longer exposure time. These additional data consisted in a set
of observations made between April 15 and May 23 2000 with a total
exposure time of 33 ks. Unfortunately,  although we found a similar
trend for negative lags at low frequencies,
 the results were still inconclusive.   

On the other hand, the data enable us to clearly 
detect the optical light curve lagging the X-ray one. The optical
phase lags as
 a function of Fourier frequency are displayed in Fig.~\ref{fig:corxofour}c.
These phase-lags increase
significantly with the Fourier frequency covering a broad range of
angles centered around $\pi/2$.
 This leads to time lags (shown
in Fig.~\ref{fig:lago})
decreasing with frequency approximately like $f^{-0.8}$ in the
 range 10$^{-2}$--$10$ Hz and spanning roughly 10$^{-2}$--10 s 
over this frequency range. 
The frequency dependence of the lags is not exactly a power law as
 additional structures are apparent, in particular around 0.5 Hz.
Such structures are even more apparent in the phase lag diagram.
Interestingly, as in the case of X-ray lags the data suggest the
presence of negative lags at low frequencies. The significance of 
these negative lags is however very low.

The optical vs X-ray coherence  function (shown in Fig.~\ref{fig:corxofour}b) is frequency dependent 
as well. It reaches a maximum in the 0.1--1 Hz range and decreases rapidly
 both at lower and higher frequency. 
The overall coherence is not very high ($<0.3$) indicating that only 
a relatively small fraction of the optical and X-ray variability are
 actually correlated. As the coherence gives 
only the \emph{relative} importance of correlated variability, it is
not clear whether the observed frequency dependence is a property of
of the physical process producing the correlation, or if it   
is driven mainly by changes in the amplitude of the uncorrelated 
fraction of the variability. We note that part of the loss of coherence
around 10 Hz is due to the artifact QPO visible in the optical power
spectrum that is  due to telescope oscillations (see Spruit \&
Kanbach 2002).

This Fourier analysis is consistent with the CCF shape.
Basically the CCF plots the coherence as a function of time-lags
 weighted with the fractional amplitudes of variability 
at the frequencies producing such 
lags. The CCF indeed appears to peak at values $\sim 0.3$ at
 lags $\sim 0.5$ s i.e. roughly corresponding to the frequency range 
where coherence is maximum.
  
In addition, the reduction of the optical time-lags  with frequency while 
the phase-lags increase is qualitatively consistent
 with the changes of the optical peak 
lag in the CCF at different time-scales of the fluctuations (shown in
Fig.~\ref{fig:ccfscadep} and discussed in Sect.~\ref{sec:oxcor}).
On the other hand the Fourier time-lag diagram does not show any evidence for
 the optical dip at negative lags apparent in the optical X-ray CCF.
Indeed, since in the Fourier method, the light curves are decomposed
 over cosine functions, an anti-correlation at negative optical lags appears
 exactly in the same way as a positive correlation at positive lags. Thus,
 the optical dip and peak information are all mixed together in the Fourier 
representation.

\section{Superposed shot and dip analysis}\label{sec:superposition}

%                                     Two column figure (place early!)
%______________________________________________ Gamma_1 (lg rho, lg e)

   \begin{figure*}
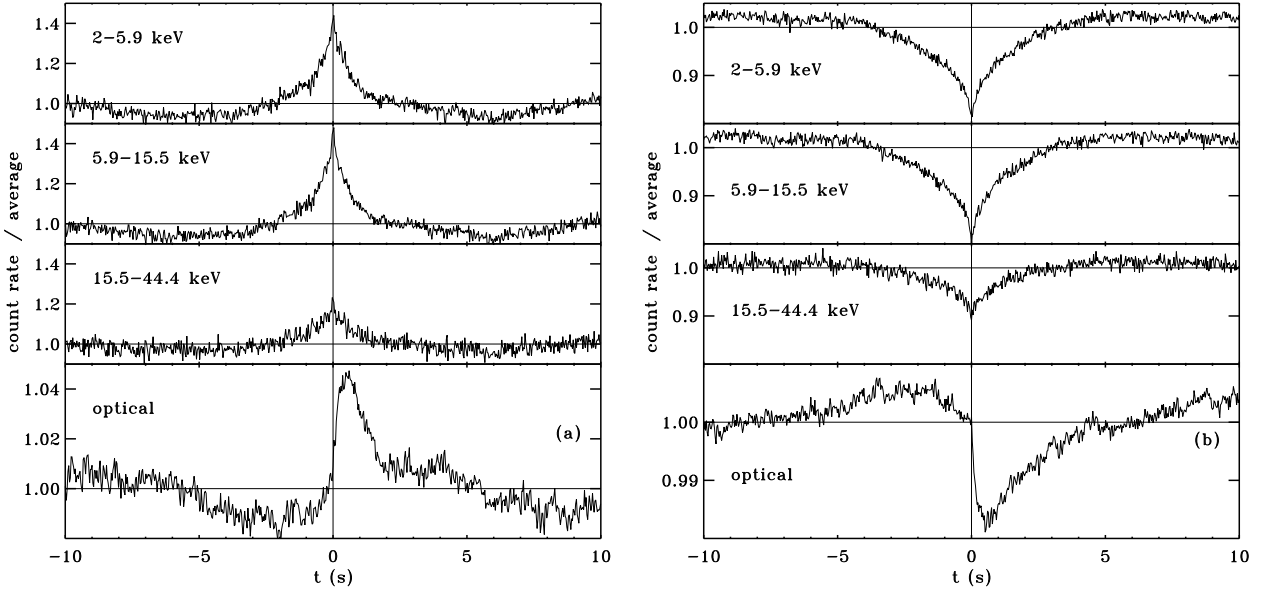

   \centering
   \scalebox{0.4}{\includegraphics{3498.f9a}}\hspace*{0.5cm}\scalebox{0.4}{\includegraphics{3498.f9b}}

   \caption{Results from the shot (panel {\bf a}) and dip (panel {\bf b})
superposition technique when the flares or dips are
selected in band b1. The selection parameters (see
Sec.~\ref{sec:superposition}) are $f$=2, $t_{p}$=8 s,
$t_{m}$=32 s in both panels.}
              \label{fig:shotdipx}
    \end{figure*}

   \begin{figure*}
   \centering
   \scalebox{0.4}{\includegraphics{3498.f10a}}\hspace*{0.5cm}\scalebox{0.4}{\includegraphics{3498.f10b}}

   \caption{Results from the shot (panel {\bf a})  and dip (panel
{\bf b})
superposition technique when the flares or dips are
selected in the optical band. The selection parameters  (see
Sec.~\ref{sec:superposition}) are $f$=1.2, $t_{p}$=8 s
$t_{m}$=32 s in both panels}
              \label{fig:shotdipopt}
    \end{figure*}
   \begin{figure*}
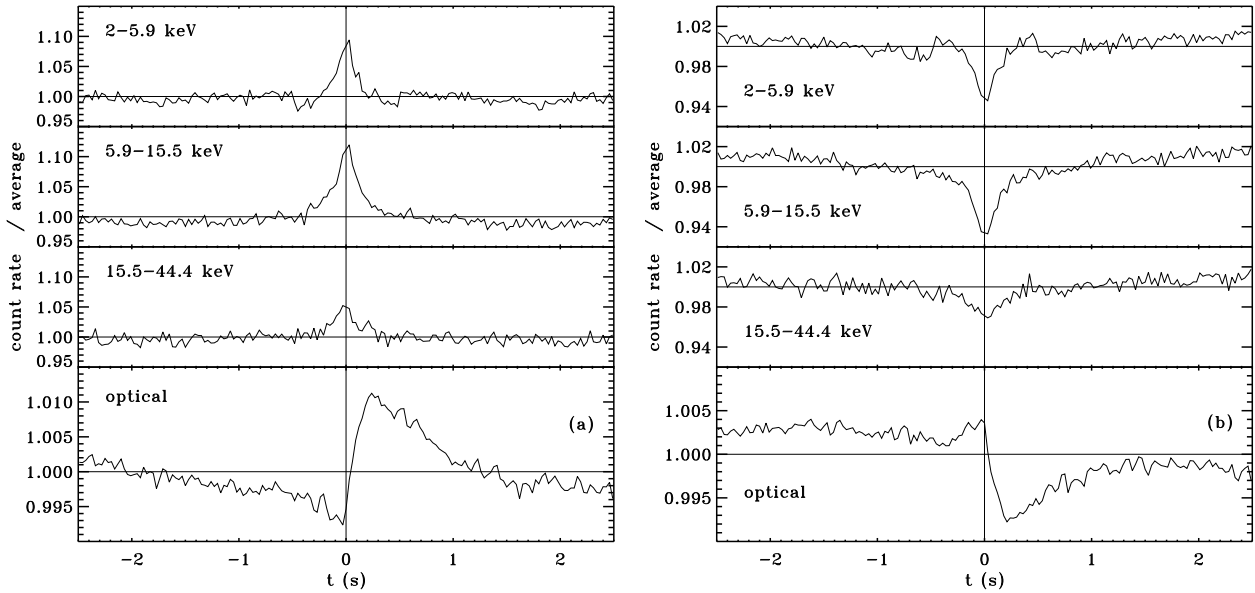

   \centering
   \scalebox{0.4}{\includegraphics{3498.f11a}}\hspace*{0.5cm}\scalebox{0.4}{\includegraphics{3498.f11b}}
   \caption{Results from the shot (panel {\bf a}) and dip (panel
{\bf b})
superposition technique using band b1 as the selecting band.
 The selection parameters  (see
Sec.~\ref{sec:superposition}) are $f$=1.8, $t_{p}$=0.5 s and $t_{m}$=5 s.}
              \label{fig:shotdipsmall}
    \end{figure*}

Besides Fourier and time domain analysis, there is a third method
 occasionally used to study the variability of X-ray binaries. 
It is based on the selection and averaging of flares events 
in the light curves.  
 
In the following we will apply a method similar to that used by
Negoro, Kitamoto \& Mineshige  (\cite{negoro2001}) to study the
average shot profile in Cyg X-1.
We select flare events in the lower energy light curve (b1, 2--5.9 keV).
These shots are selected according to the same criteria as those used by
 Negoro, Kitamoto \& Mineshige  (\cite{negoro2001}): 
the peak count rate of the shot is $f$ times the local count rate as obtained 
from an average over time $t_{m}$. 
The peak bin is further required to have the maximum count-rate over bins 
within $t_{p}$ before and after the peak bin.
The selected shots are then peak aligned and averaged. 
The corresponding pieces of light curves in the b2, b3 and optical band
 are centered on the time bin corresponding to the b1 peak and averaged
 in the same way.

Fig.~\ref{fig:shotdipx}a shows the results for $f$$=$2, $t_{p}$$=$8 s 
and $t_{m}$$=$32 s. The light curves were rebinned on 30 ms time 
bins before applying the shot selection.  
For such parameters the resulting average X-ray shot is slightly asymetric,
with a duration of $\sim 10$ s. The b2 and b3 bands appears to present 
shots that are similar to that in the b1 band. This illustrates the high
 degree of coherence between the different energy bands. 

The shot in
the b3 band has a lower amplitude than the b1 and b2 shots.
This is consistent with the energy dependent power spectrum indicating a 
lower amplitude of variability in the higher energy band. In addition,
 this enables us to see in a more direct way the spectral evolution
 leading to the anti-correlation between X-ray hardness and flux 
discussed in Sect.~\ref{sec:tdanal}.
The average optical light curve corresponding to the shots is 
similar in shape to that of the optical/ X-ray CCF. This suggests,
 as previously noted
by Spruit and Kanbach (\cite{spruit2002}), that the shape of the CCF
 is representative of the shape of the optical light curve as a
response to a shot event.

Then it is interesting to see whether the optical light curve responds
only to X-ray flares or is also correlated to other types of
events occuring in the X-ray light curve. 
We thus performed a similar analysis but instead of flares, we 
selected dips in the b1 band. The selecting criteria were that the
minimum count rate of the dip is lower than $1/f$ times the local 
count rate as obtained from an average over $t_{m}$.
The minimum bin is further required to have the minimum count-rate
 over the bins within $t_{p}$ on either side.

The results are shown in Fig.~\ref{fig:shotdipx}b.
Surprisingly, the answer is that the optical responds to X-ray dips
 in a similar but inverted way as it responds to the X-ray flares.
The optical flux rises a few seconds before the minimum in the X-ray
light curve, at t=0 it decays abruptly with a minimum  half a second 
after the X-ray dip. The panels (a) and (b) 
of Fig.~\ref{fig:shotdipx} are actually very similar but with
inverted count-rate axis.

To learn about  the response of the X-rays to optical fluctuations 
 we performed a similar shot and dip analysis using the
optical band as the selecting light curve. The results are shown in 
Fig.~\ref{fig:shotdipopt}. The X-ray shots and dips are not simply the
same as in Fig.~\ref{fig:shotdipx} shifted by $\sim$ 0.5 s as one
would expect if the correlation was linked only to the X-ray dips and shots.
The X-ray response to optical shots and dips
 is very asymmetric. The optical
shots are associated with an X-ray flux that
rises slowly during a few seconds and decays sharply in $\sim$ $0.5$ s.
There is an indication for the presence of 
an X-ray dip after the peak in the optical.
A similar asymmetry is apparent in the dip analysis.
Thus from  Figs.~\ref{fig:shotdipx}~and~\ref{fig:shotdipopt}, it
 appears that the correlated  optical and X-ray shots and dips all
have a profile resembling that of the optical/X-ray CCF (modulo the
relevant symmetries).

The superposition method is far less rigorous
 than usual time-domain and Fourier techniques.
 In particular the results are affected by strong biases. 
It tends to favor a certain range of time-scales
 and amplitude depending on the selection criteria.
 In general, the selected events are not representative
 of the whole variability of the source.
It could be also, that the light curves are not
made of a superposition of shots at all.  
Moreover a description of the variability of accreting black hole
sources in terms of shots models (e.g. Poutanen \& Fabian 1999 and
reference therein) requires, in general, 
events with a broad range of time-scales and amplitude.
The superposed events are not representative 
of a clearly defined scale, rather, they represent 
an average over a range of scales that is poorly controlled.

The drawbacks of the method can be taken into our advantage. By
 changing the selection criteria one can select (in a very qualitative way)
 the time-scale and amplitude of the typical selected events. 
For instance, Fig.~\ref{fig:shotdipsmall}
 shows the results of the shot and dip 
superposition for $t_{p}$$=$0.5 s and $t_{m}$$=$5 s and $f$$=$1.8
selected in the b1 band.
 The resulting average 
profiles have clearly a lower amplitude and shorter duration ($< 1$) s.
The interesting thing is that the both the X-ray and optical light curves
 look like a rescaled version of Fig.~\ref{fig:shotdipx}. 
In particular, the softening of the spectrum with the enhanced count rate,
and the presence of an optical dip and flare respectively before and after
 the X-ray peak.

 The superposition analysis thus confirms that
  both the optical pre-dip and post-peak are present over a wide range of 
time-scales. 

More importantly, the superposition method,
despite all its biases, 
reveals that the optical/X-ray correlation is not triggered by a
single type of event (flare or dip) in the light curves.
The correlation is thus a global property of the light curves.
Flares and dips of different shapes and  time-scales
appear to contribute to the shape of the total optical X-ray CCF.
Strikingly, they 
all contribute in a similar way to the formation of 
the pre-dip and post-peak features.

Also, when comparing the superposed light curves
we see that the changes occur on similar time-scales, and almost 
at the same time in the different bands.
The general impression is 
that high (low) optical fluxes are associated with a sharp decay
(rise) in the X-rays. 
 In other words the optical appears to be proportional to the opposite of
the time derivative of the X-rays.
If such a relation was to hold strictly, we would expect a constant
optical phase-lag of value $\pi/2$ (due to the fact that a derivation
in the time domain translates into a constant phase 
factor in the Fourier domain). 
In contrast, we found (see Fig.~\ref{fig:corxofour}) that the
phase-lags do depend on Fourier frequency and approximatively cover the range
0--$\pi$.
Nevertheless, we see that the phase lags are close to $\pi/2$ 
in the frequency
range 0.1--1 Hz where the coherence is significant.
 Therefore, in the frequency range where most of the
correlation occurs, the optical is indeed related to the time
derivative of the correlated X-ray flux.
Equivalently, the $\sim \pi/2$ lags can be interpreted as the X-ray light
curve scaling like the time derivative of 
its correlated optical counterpart.
 Due to the remarkable symmetry of the correlation, this interpretation would be also roughly consistent with the general trend
observed in the superposition
analysis (Fig.~\ref{fig:shotdipx},\ref{fig:shotdipopt},
\ref{fig:shotdipsmall}).
There is thus a 'differential' relation between the optical and the
X-rays: the variability of a light curve
appears related to the time derivative of the correlated fraction 
of the other.

With this regard, rather than using concepts
like `precognition' dip and delays,
 the shape of the superposed light curves and CCF  
could be interpreted as the signature of a process 
affecting at the same time, both the X-ray and optical emitting
region.
In this context, it is not clear whether the 'differential' relation
reflects an intrinsic 'differential' property of the underlying
physical process, or on the contrary, emerges casually from the 
fact that the phase-lags are
close to $\pi/2$ in the frequency range where coherence is important.

\section{Discussion} \label{sec:discussion}

Several studies already emphasized the similarity of the X-ray power spectrum
 in XTE~J1118+480
with that of typical black hole candidates in the hard state.
 Here we find that the similarity extends to other X-ray timing
features, namely:
\begin{itemize}
\item the weak energy dependence of the power spectrum, 
\item the reduction of the integrated rms amplitude at high energy,
\item the shape of the X-ray ACF and CCFs,
\item the presence of an anti-correlation between the X-ray flux and hardness,
\item the unity coherence between X-ray bands over a wide range of frequencies.
\end{itemize}

Together with the presence of striking spectral similarities 
(Frontera et al. \cite{frontera2001}, \cite{frontera2003}), this probably makes XTE J1118+480 
a typical hard state source.
 On the other hand our study suggests that
XTE J1118+480  might differ from typical hard states sources by the
presence of soft lags at frequencies below 0.1 Hz apparent both in the 
Fourier analysis and through the asymmetry found in the X-ray CCFs.

Our study of the correlated optical/X-ray flickering reveals that:

\begin{itemize}
\item the optical X-ray CCF is independent of the X-ray band.
\item the optical band is correlated not only to the X-ray flux but also,
and in a similar way, to the rapid X-ray spectral fluctuations. This
is shown by the computation of the optical flux vs X-ray 
hardness CCF and the superposition analysis.
\item the optical phase-lags increase
significantly with Fourier frequency covering the range 0--$\pi$.
 This leads 
to time-lags decreasing roughly like $f^{-0.8}$.
\item the optical/X-ray coherence is maximum ($\sim0.3$) for fluctuations 
       in the range 1-10 s and decreases sharply both at shorter and
longer time-scales. 

\item the shape of the X-ray CCFs asymmetry
(Fig.~\ref{fig:xxccfs}) is 
reminiscent of the asymmetry of the optical/X-ray CCF 
(Fig.~\ref{fig:ccfox}a). This may indicate a connection between the X-ray
and optical lags. This suggestion is strengthened by the dependence of the
optical time-lags at higher frequency $\sim f^{-0.8}$ that is
comparable to the dependence of the X-ray lags in hard state black holes.
\item the filtered CCF analysis as well as the superposition technique
      shows that the underlying process, giving rise to the correlated
variability, acts in a similar but rescaled way at different time- and amplitude-scales of the fluctuations.
\item the superposition technique shows that the optical response
is not triggered by a specific type of event in the X-ray light curve (shot
or dip). Rather, events of very different shapes and time-scales
contribute in a similar way to the shape of the optical/X CCF.
In the range of frequencies where the optical/X-ray coherence is maximum,
 the light curve in one band is related to the time
derivative of its correlated counterpart.

\end{itemize}

One of the main result of this work 
is that the optical/X-ray correlation can be 
 described as self-similar. 
  We note that this scaling in time is consistent 
with what SK02 find by the completely independent method of
principal component analysis (PCA).  With this analysis, the
X-ray/optical
 cross correlation is computed on short segments of data (25s long),
 yielding a large number of
samples (300) of the cross correlation, which turn out to somewhat
 variable in shape. With the PCA (e.g.\ Kendall, 1980), the similarities
and difference between samples are analyzed by finding linear
combinations
 of the samples which are statistically uncorrelated. If the
variability of the signal comes about as a mixture of a few signals
 of fixed shape but independently varying amplitudes, the PCA will
extract these shapes as the `principal components'. In SK02, two 
significant components were found, both of which were of the
dip-plus-spike shape but on time scales differing by a factor 3. 
This is similar to the results presented here (compare Fig 9 lower
left with Fig 11 lower left panel). The `scaling' of the components 
suggests that there is just one underlying process, but that it can
take place on different time scales, possibly in a continuous fashion.
The filtered CCF analysis (see Fig.~\ref{fig:ccfscadep}) and the
power-law dependence of optical time-lag on Fourier frequency 
(Fig.~\ref{fig:lago}), suggest 
that indeed, this underlying process is continuously self-similar.
Actually, the PCA is not well 
suited to study such a continuous variations. 
If the signal shape varies in a continuous way, the PCA will
represent this continuum by a number of discrete components.
 These components will be representative of the range of variation
through this continuum. Moreover, the PCA is sensitive to the coherence and 
amplitude of variability of the two signals at a given time-scale.
Only the time-scales where both the coherence and the amplitude of
variability is high will contribute significantly to the variability
of the CCF.
The finding of SK02 that there are only
 two significant components simply
reflects the fact that the coherence is high only in a relatively 
narrow range of Fourier frequencies. In other words, 
the \emph{shape} of CCF at different time-scales appears 
to be continuously self-similar, but 
the \emph{amplitude} of the average CCF 
is dominated by the two time-scales evidenced 
by the PCA of SK02 and our shot selection analysis.

This scaling is not readily explained by the current accretion models. 
The ideas that were proposed so far to explain 
the long optical lags were based on the propagation time 
between two distinct X-ray and optical emitting regions
 that are causally connected.
In the reprocessing models the lags are interpreted as 
the light travel-time between the corona and the disc.
In the Kanbach et al. \cite{kanbach2001} scenario a
 perturbation propagates
 from the X-ray corona to the optical outflow photo-sphere.
In such propagation models,
 the nearly scale-invariant correlation then requires
 that the small-scale fluctuations have a large propagation
 speed (outflow model) or alternatively that there are numerous pairs
 of X-ray/optical emitting regions at different distances from each other,
 the small-scale perturbation being produced in the closest pairs.
Propagation models thus appear strongly constrained 
and require some complications to fit the scale-invariance. 
Also as discussed in K01 the propagations models encounter important
difficulties in producing features such as a precognition dip that 
occurs together with a post-peak.

Although the lags 
indicate a form of causal connection
between the X-rays and optical emitting regions they 
do not necessarily involve a propagation process.   
Rather, the event superposition analysis suggests that the complicated 
correlation could be the signature for
 a self-similar process acting \emph{simultaneously} in both
emitting regions with different manifestations in both bands.

 The apparent advances/delays of the optical signal with respect 
to the X-rays 
could then be a consequence of a (somewhat complex) evolution of 
the spectral
shape of the emitted radiation during an X-ray event. Such an 
explanation
has already been proposed for the lags observed between hard
and soft X-rays in Cyg X-1 (e.g. Poutanen \& Fabian 1999).
In this 
case the hard and soft X-rays are be produced by a single emission
mechanism and in the same physical volume. 
In our case however the situation is more complex.
 The optical/X CCF is much more complex than the X-ray CCFs and
 the emission process in the two band is probably different.
It is also not clear whether the X-ray and optical emission are
produced in the same physical region.
Finding a consistent physical scenario that would
make such a model predictive and testable is very challenging.

In the jet model of Markoff, Falcke \& Fender (\cite{markoff2001}),
the X-rays and optical are both 
formed through synchrotron emission in the jet. The X-ray CCF could
be, in principle due to a characteristic time-evolution of 
 of the distributions of the emitting particles in the jet.
The difficulty, here, is that no obvious physical mechanism 
can produce such a spectral evolution.

In other models, such as the ADAF model of Esin et al. (\cite{esin2001})
or  the coronal model of  Merloni, Di Matteo \& Fabian
(\cite{merloni2000}) both the optical and X-rays are emitted in the
hot inner part of the accretion disc. 
For instance, if 
 we assume that the optical is self-absorbed synchrotron
radiation produced by the Comptonising cloud itself (with a temperature
$kT_{e}\sim 100$ keV as estimated by Frontera et al. \cite{frontera2001}, \cite{frontera2003}).
 Then, the observed optical
flux implies a size of the emitting region of a few hundred
Schwarzschild radii. Incidentally, this dimension  
is of the same order as the 
cold disc inner radius estimated in XTE~J1118+480 from fits of the optical to EUV
spectra (see Chaty et al. 2003). Then the correlated variability could
be caused, for instance,
 by temporal fluctuations of the magnetic field
and/or optical depth of the comptonising plasma.
The details are however to be precised and physically motivated.
 This scenario also encounter
 a problem to reproduce the observed optical spectrum.
 This spectrum indeed appears much flatter than
that produced through self-absorbed synchrotron.
 Thus, again, some complications
 (e.g. strong gradients of magnetic field and/or
plasma temperature of the Comptonising plasma) would be required
to fit the spectral data.
 
It could be also that,
 contrary to what is assumed in the above mentioned models,
 the X-ray and optical are produced in two
distinct regions  (e.g. disc+jet). In this case a spectral evolution
scenario could also, in principle, 
account for the correlated X-ray optical variability.
Then, the two regions should be 
physically connected on a time-scale shorter
 than $\sim$ 0.1 s, that is the shortest time-scale on which 
we have been able to observe the typical spectral evolution.
The complication with such a model is that it 
should provide a physical description of the connection between the
two regions that is well beyond the simple radiation physics.
In the case where the X-rays come from the disc and the optical from
the jet, the optical X-ray CCFs, their self-similarity
and the apparent 'differential' relation between the two light curves
would constitute important signatures of  
the underlying accretion-ejection process.

\begin{acknowledgements}
 This work was partly supported by the European Commission
 (contract number ERBFMRX-CT98-0195, TMR network
"Accretion onto black holes, compact stars and protostars").
JM also acknowledges fundings from the MURST (COFIN98-02-15-41) and
PPARC. We thank Andrea Merloni for valuable discussion.
\end{acknowledgements}

\end{document}